\documentstyle[prl,aps,multicol,epsf,epsfig,rotate]{revtex}
\newcommand \be {\begin{equation}}
\newcommand \ee {\end{equation}}
 \newcommand \ba {\begin{eqnarray}}
\newcommand \ea {\end{eqnarray}}                        
\begin{document}
%\draft
\tightenlines

\title{\bf Population dynamics in a random environment}
\author{Irene Giardina$^*$, Jean-Philippe Bouchaud$^\dagger$, Marc M\'ezard
$^{**}$}

\address{$^*$ Service de Physique Th\'eorique et $^\dagger$ Service de Physique de l'Etat Condens\'e}
\address{Commissariat \`a l'Energie Atomique, Orme des Merisiers}
\address{91191 Gif-sur-Yvette {\sc cedex}, France}

\address{$^{**}$ Laboratoire de Physique Th\'eorique et Mod\`eles Statistiques
}
\address{Universit\'e Paris Sud, Bat. 100, 91 405 Orsay {\sc cedex}, France}
\date{\today}
\maketitle

\begin{abstract}
We investigate the competition between barrier slowing down and proliferation 
induced 
superdiffusion in a model of population dynamics in a random force field. 
Numerical results in $d=1$ suggest that a new intermediate diffusion behaviour appears. We introduce 
the idea of proliferation assisted barrier crossing and give a Flory like 
argument to understand 
qualitatively this non trivial diffusive behaviour. A 
one-loop {\sc rg} analysis close to the critical dimension $d_c=2$ confirms that the random force fixed point is unstable and flows towards an uncontrolled strong coupling regime. 
\end{abstract}
\begin{multicols}{2} 

\narrowtext

The presence of disorder often radically changes the statistical properties 
of random walks. For example, random walks in a random potential are trapped 
in deep potential wells: this may lead to
{\it sub-diffusion}, i.e. the fact that the typical distance traveled
by the walkers grows more slowly than the square-root of time \cite{BG}. A 
much studied model exhibiting this type of behaviour is the Sinai model, 
where particles diffuse in a random force field in one dimension 
\cite{Sinai,Annals,FLM}. In this case, the energy barriers typically grow as 
the square root of the
distance, which leads to a logarithmically slow progression of the random 
walkers. There are also several mechanisms which lead to 
{\it super-diffusion}. For example, if the random force field is 
rotational, the random walkers can be convected far away by long streamlines 
\cite{BG}. Another interesting mechanism of superdiffusion is 
{\it random proliferation}: suppose that each random walker can 
either die or give birth to new random walkers at a rate which is random, 
both in time and space. There is in this case a possibility for an `outlier' 
random walker, that has by chance traveled a distance 
much greater than the square-root of time, to have been particularly 
prolific: he and his siblings then represent an appreciable fraction 
of the whole population, leading to a motion of the center of mass 
faster than diffusive. This mechanism has been much studied (although not 
explicitely discussed as such) in the context of Directed Polymers ({\sc dp}) in
random media or equivalently the Kardar-Parisi-Zhang ({\sc kpz}) model of surface 
growth \cite{HHZ}. The aim of the present work is to investigate the case 
where both these mechanisms are present simultaneously. The motivations for 
such a mixed model are numerous. In the context of 
population dynamics (for example, bacteria on a random substrate), similar 
models have recently been investigated, with quite interesting results 
\cite{Nelson}. One can also give an economic interpretation of population 
dynamics, where the local density of random walkers is the wealth of a given 
individual. Biased diffusion represents trading between individuals, whereas 
the random growth term is the result of speculation \cite{BM}. One can argue 
that generically, this type of model leads to a Pareto (power-law) tail in 
the distribution of wealth \cite{BM}. Finally, from a theoretical point of 
view, this mixed model leads to the interesting possibility of new behaviour,
intermediate between superdiffusion and subdiffusion.

More precisely, we study here the following equation for the local 
population density $P(\vec x,t)$ in $d$ dimensions:
\be
\frac{\partial P(\vec x,t)}{\partial t}= \nu_0 \Delta P(\vec x,t) -
\vec \nabla (\vec F(\vec x) P) + \eta(\vec x,t) P(\vec 
x,t),\label{mixed}
\ee
where $\nu_0$ is the bare diffusion constant, $\vec F(\vec x)$ a space
dependent static Gaussian random force such that $\langle F_\mu(\vec x)
F_\nu(\vec x')\rangle_F = 2\sigma_F^2 \delta_{\mu,\nu} \delta^d(\vec x - \vec 
x')$ \cite{Rq0}, and $\eta(\vec x,t)$ a Gaussian random growth rate, depending both on space and 
time, with $\langle \eta(\vec x,t)
\eta(\vec x',t')\rangle_\eta = 2\sigma_\eta^2 \delta(t-t') \delta^d(\vec x - 
\vec x')$. Due to the last term, the total
population $Z(t)=\int d\vec x P(\vec x,t)$ is not conserved. The quantities 
of interest, which describe how the population spreads in time are, for 
example, the average center of mass motion,
\be
x_{cm}^2(t)=\langle \left(\frac{1}{Z} \int \vec x P(\vec 
x,t)\ d\vec x\right)^2   \rangle_{F,\eta}
\ee
or the average `width' of the diffusing packet $\langle \Delta^2 
\rangle_{F,\eta}$ :
\be
\Delta^2(t)= \frac{1}{Z} \int  \vec x^2 P(\vec x,t)\ d\vec x - 
\left(\frac{1}{Z} \int \vec x P(\vec x,t)\ d\vec x \right)^2 
\ee
(Other moments can however also be studied: see below). An alternative 
description is in terms of the free-energy $h(\vec x,t)=
\log P(\vec x,t)$, which obeys 
the equation:
\begin{eqnarray}\nonumber
\frac{\partial h(\vec x,t)}{\partial t}&=& \nu_0 \Delta h(\vec x,t) 
+ \lambda (\vec \nabla h)^2 \\ &-& \vec F(\vec x)\cdot \vec \nabla h
- \vec \nabla \cdot \vec F(\vec x) + \eta(\vec x,t),\label{kpz}
\end{eqnarray}
with $\lambda=\nu_0$. When $\vec F \equiv 0$, these equations represent the well-known {\sc kpz} (or 
Directed Polymer) problem, whereas for $\eta \equiv 0$, one recovers the 
problem of a random walk in a random environment. Both problems can be 
approached using a perturbative Renormalization group;
interestingly, the critical dimension for both problems is $d_c=2$. For the 
random drift problem,
one finds that the coupling constant $g_F=\sigma_F^2/(2\pi)\nu_0^2$ flows towards a 
non trivial fixed point of order $\epsilon$ in dimensions $d=2-\epsilon$ 
\cite{Luck,Fisher,BG}. This in turn leads to a subdiffusive behaviour: 
$x_{cm}(t)$ grows as $t^\nu_F$ with $\nu_F=(1-\epsilon^2)/2 < 1/2$. 

\begin{figure}
\hspace*{-0.5cm}\epsfig{file=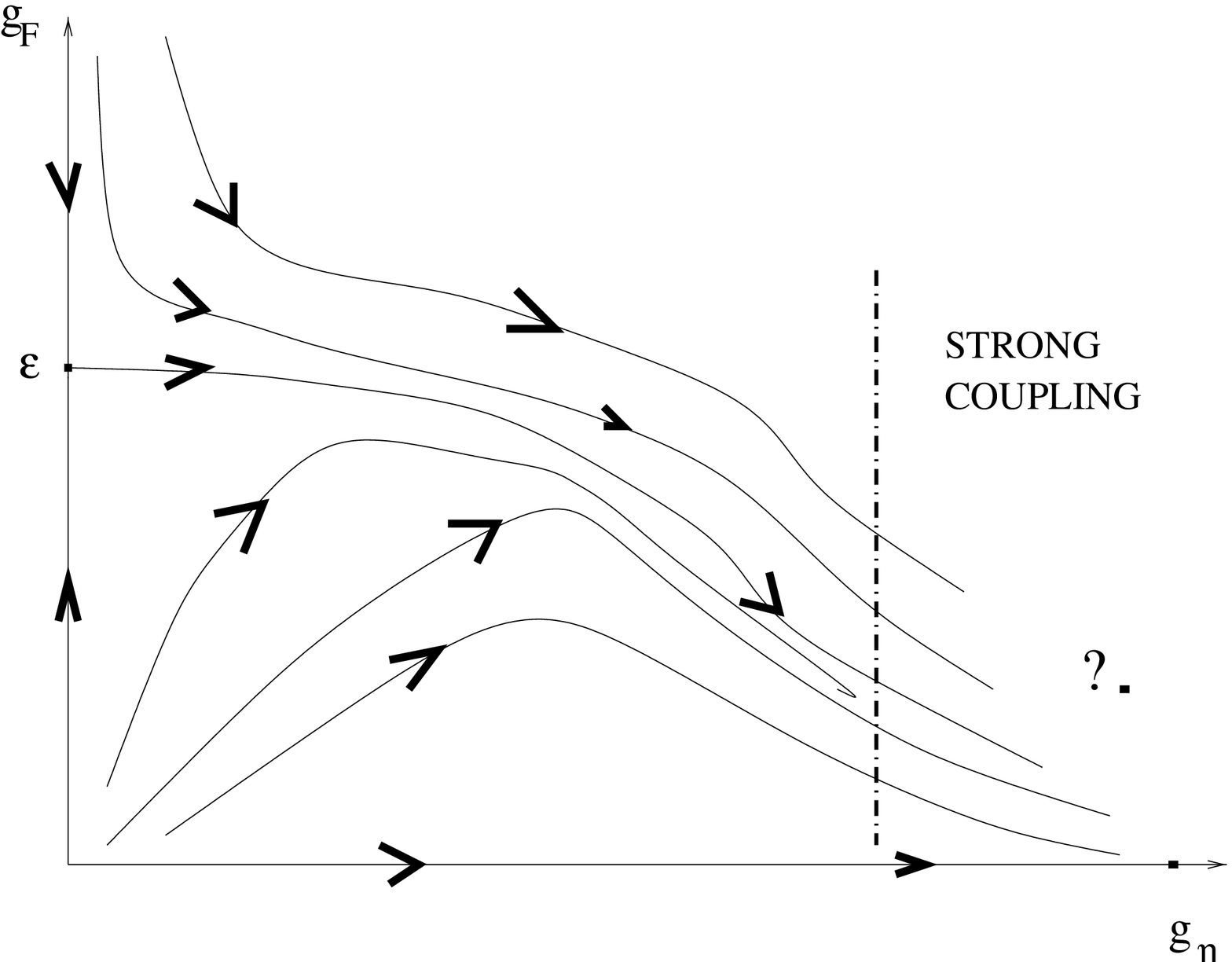,width=7cm}
\caption{\small One loop {\sc rg} flow in the $g_\eta, g_F$ plane. As soon as $g_\eta$ is non zero, it flows towards the strong coupling region. Our numerical simulations suggest that an attractive `mixed' fixed point appears in this region.\label{fig1} }
\end{figure}
For the 
{\sc kpz} problem, the coupling constant is $g_\eta=\sigma_\eta^2 
\lambda^2/(2\pi)\nu_0^3$; the Gaussian fixed point 
$g_\eta=0$ is again unstable for $d<2$, but there is no accessible fixed 
point at one loop for $d>3/2$ \cite{RGKPZ}. The exponent $\nu$ therefore 
cannot be computed but is expected (and found numerically) to be greater than 
$1/2$: in the population dynamics language, the possibility of far-away 
proliferation leads to
superdiffusion. On the other hand, for $d<3/2$, a physical fixed point 
appears; in $d=1$, the one loop calculation even provides the exact result 
$\nu_\eta=2/3$ for reasons detailed in \cite{RGKPZ}. We have 
performed a {\sc rg} analysis in the {\it mixed} case where both $g_F$ and $g_\eta$ 
are non zero. This can be done using a field theoretical representation 
(Martin-Siggia-Rose) representation of Eq. (\ref{kpz}), which allows one to 
generate the perturbation expansion in $g_F$ and $g_\eta$. Performing 
calculations along the lines of \cite{Luck,Fisher,RGKPZ}, we find that the 
two $\beta$
functions are given by:
\be
\frac{dg_\eta}{d\ell}= \epsilon g_\eta+2g_\eta g_F+\frac{g_\eta^2}{4}; \frac{dg_F}{d\ell}= \epsilon g_F-\frac{\epsilon g_\eta g_F}{4}-g_F^2,
\ee 
where $\ell$ is the logarithm of the running length scale. The resulting flow 
is represented in Fig. 1. The subdiffusive random force fixed point $g_\eta=0$, $g_F=\epsilon$ is therefore unstable in the presence of a small `proliferation' term $g_\eta$. Unfortunately, at one loop, $g_\eta$ flows towards the
strong coupling region, as is the case in the standard {\sc kpz} case $g_F=0$. 

In order to obtain some information about this strong coupling behaviour, 
we have performed some numerical simulations in one dimension, where both the fixed points 
corresponding to Sinai subdiffusion and to {\sc dp/kpz} superdiffusion are 
rather well 
understood. We have found results that suggest the existence of an
{\it attractive} fixed point, characterized by a new non trivial diffusive behaviour (intermediate between the Sinai and {\sc dp/kpz} behaviour).
We have numerically 
evolved a space and time discretized version of Eq. (\ref{mixed}), and have 
worked with $\log P$ to avoid
precision problems. Starting from a localized packet 
$P(x=ia,t=0)=\delta_{i,0}$, we have found that as soon as both coupling 
constants $g_F^0$ and $g_\eta^0$ are non zero, the exponent $\nu$ describing 
the diffusion of the center of mass $x_{cm}(t)$ at large times is found to be close to the value $\nu^*=1/2$ (see Fig. 2). 
The position $x_{max}(t)$ of the maximum of $P(x,t)$ behaves very similarly.
\begin{figure}
\hspace*{-0.5cm}\epsfig{file=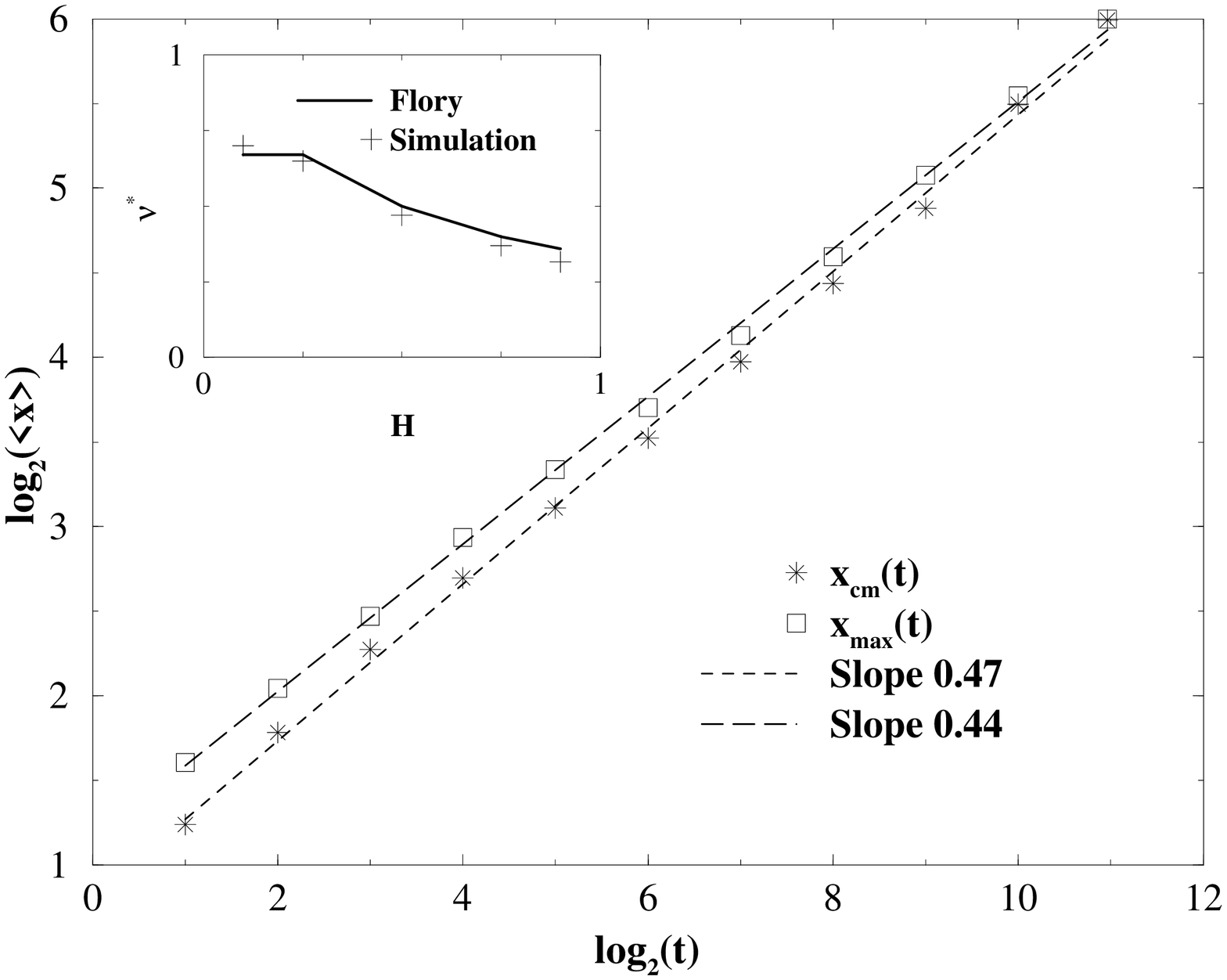,width=7cm,angle=270}
\caption{\small Behaviour of the average center of mass $x_{cm}(t)$ and of the average position of the maximum of the packet $x_{max}(t)$ (i.e. the point where $P(x,t)$ is maximum), as a function of time, for $\sigma_\eta/\sigma_F=0.125$. The best linear fits are shown, and lead to the an estimate for $\nu^*$ slightly smaller than $1/2$. Inset: Value of $\nu^*$ (determined from the behaviour of $x_{cm}(t)$) as a function of the Hurst exponent of the potential $H$, compared with the Flory prediction. \label{fig2} }
\end{figure}
 The ratio $g_F^0/g_\eta^0$ affects only the 
short time transient behaviour, which is either Sinai like or {\sc dp/kpz} like, as
shown in Fig. 3. In the {\sc rg} language, this suggests that a non trivial {\it attractive} fixed point $g_F^*,g_\eta^*$ appears. This is compatible with 
the flow diagram of Fig. 1 \cite{footnoteflow}, although this new fixed point is out of reach at the one-loop level. Although 
the value of $\nu^*=1/2$ corresponds to free-diffusion, the motion of the 
packet for a given environment is very far from simple diffusion, as the 
study of the width $\Delta$ of the packet shows. We have found numerically 
that $\langle \Delta^q
\rangle_{F,\eta}$ behaves as $t^{q\zeta_q}$ with $\zeta_1\simeq 0.24$, 
$\zeta_2\simeq 0.34$ and $\zeta_4\simeq 0.38$. This non trivial behaviour is actually 
present for both the Sinai problem and the {\sc dp/kpz} problem. This comes from the fact that for both problems, the effective free energy $h(x,t)=\log 
P(x,t)$ behaves as a random walk in $x$ space. This is trivial for the Sinai 
problem, since the potential is indeed constructed as the sum of local random 
forces. For the {\sc dp/kpz} problem, this is far less trivial and results from the 
fact that one can obtain exactly the stationary distribution of $h(x,t)$ in 
one dimension, which turns out to be the same as for the linear case 
$\lambda=0$, i.e., again a random walk in $x$ space \cite{HHZ}.  It is well known that 
for a random walk potential, the probability that two nearly degenerate 
minima are separated by a distance $\Delta$ falls off as $\Delta^{-3/2}$ for 
large $\Delta$. The qth moment of $\Delta$ 
is therefore dominated by extreme events as soon as $q>1/2$. Physically, this 
means that for most realizations of the disorder, the width $\Delta$ of the 
packet is small \cite{Golosov,Mezard}, except in rare situations where the packet is divided into 
two subpackets very distant from one another. The natural cut-off for 
$\Delta$ is of the order of $x_{cm}(t)$ itself. Therefore one obtains, for 
$q>1/2$, $\langle \Delta^q(t)
\rangle \propto [x_{cm}(t)]^{q-1/2}$. For the Sinai problem, using $x_{cm}(t) 
\propto \log^2(t)$, this leads to $\langle \Delta^2(t) \rangle_F \propto 
\log^3(t)$, whereas for the {\sc dp/kpz} case, using 
$x_{cm}(t) \propto t^{2/3}$, one finds $\langle \Delta^2(t) \rangle_\eta 
\propto t$: both these results are actually exact, as has been shown in 
\cite{FLM,Mezard,FH}. Assuming that the effective potential in the mixed case
 is again a random walk in $x$ space, and using $\nu^* \simeq 1/2$,  we obtain 
$\zeta_q=(2q-1)/4q$, i.e. $\zeta_1\simeq 0.25$, $\zeta_2\simeq 0.375$ and 
$\zeta_4\simeq 0.4375$, in reasonable agreement with our numerical values 
\cite{Rq}. 
\begin{figure}
\hspace*{-0.5cm}\epsfig{file=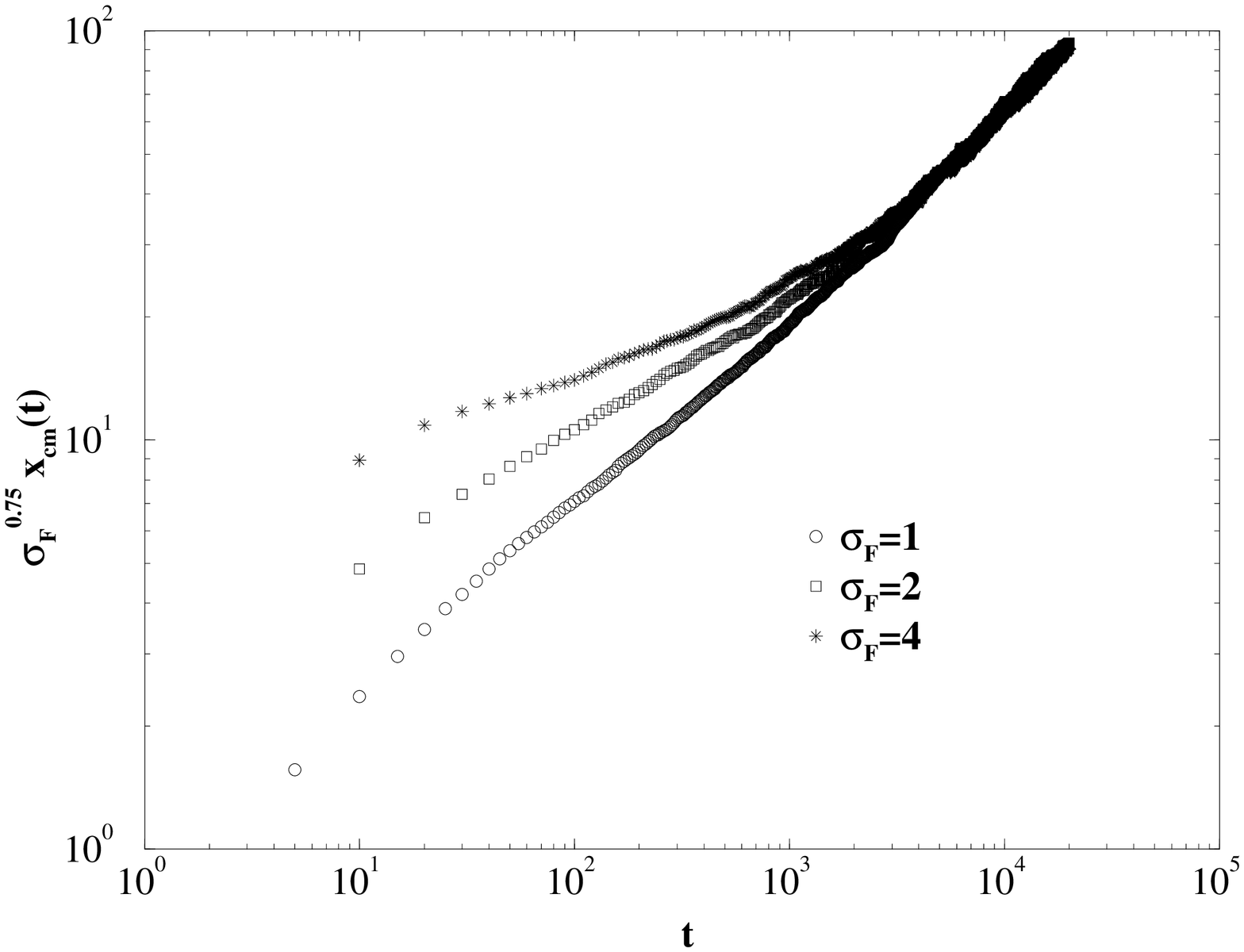,width=7cm,angle=270}
\caption{\small Behaviour of the average center of mass $x_{cm}(t)$ for
different values of $\sigma_F$. This curves shows that for large $\sigma_F$, the short time behaviour is Sinai like, crossing over to the mixed behaviour at large times. The value of $x_{cm}(t)$ has been rescaled by $\sigma_F^{3/4}$ to obtain a reasonable data collapse at large times.\label{fig3} }
\end{figure}
In order to understand the value of $\nu^* \simeq 1/2$, one needs to develop 
a consistent picture of the competition between the slowing down induced by 
the ever-growing Sinai barriers and the
speeding up of the population spreading allowed by the multiplicative growth 
term $\eta$. Before addressing the full Sinai+{\sc kpz} problem, we first consider 
the simpler case of a unique barrier
of height $U_0$, which develops on scale $L$. For definiteness, we have 
solved numerically the equation (\ref{mixed}) on the interval $[0,L]$ with 
$F(x)=-U_0/L \sin(4\pi x/L)$. The initial condition is localized in the first 
well, and the crossing time $\tau$ is defined as the average time after which 
the relative weight of the population in the second well is half of that in 
the first well. For $\eta \equiv 0$, one finds the classical Arrhenius law: 
$\log \tau = U_0/\nu_0$. When $\eta \neq 0$, the behaviour of $\tau$ as a 
function of $U_0$ for different values of $L$ is shown in Fig. 4. The result 
can be expressed as: $\tau \propto L^{3/2} f(U_0/\sqrt{L})$, with $f(y \to 
0)=1$ and
$f(y \to \infty) \propto y^b$. This scaling of $\tau$ with $L$ can easily be 
understood. In the limit $U_0 \to 0$, the time for the particles to reach a 
distance $L$ is given by the {\sc dp/kpz} scaling, i.e. $L \propto \tau^{2/3}$.

\begin{figure}
\hspace*{-0.5cm}\epsfig{file=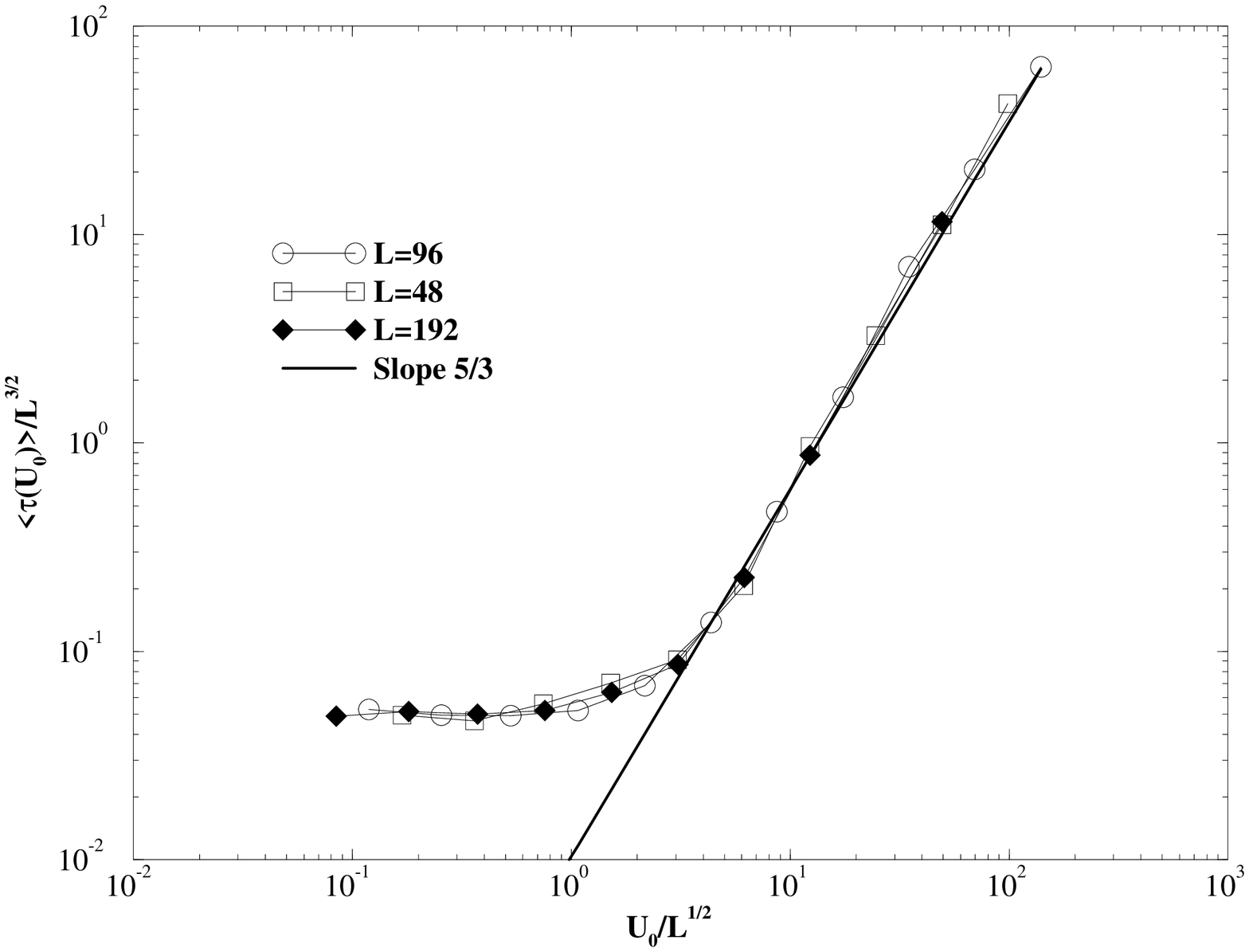,width=7cm,angle=270}
\caption{\small Average barrier crossing time $\langle \tau \rangle$, rescaled
by $L^{3/2}$, as a function of the barrier height $U_0$ rescaled by $\protect\sqrt{L}$, for different sizes $L$, and in log-log coordinates. The slope $5/3$ is shown for comparison. The power law increase of $\langle \tau \rangle$ as a function of $U_0$ has to be compared to the usual exponential (activated) dependence.\label{fig4} }
\end{figure}

 The 
influence of the external potential $U_0$ becomes substantial when it becomes 
of the order of the effective {\sc kpz} potential $h$, which, as discussed above, 
grows as $\sqrt{L}$. Numerically, the exponent $b$ is found to be very close 
to $b=5/3$. Therefore, the exponential increase of the crossing time with the 
barrier height is replaced by a power-law increase in the presence of the 
random growth term $\eta$. One can call
this effect proliferation assisted barrier crossing: the probability that a 
particle reaches the 
top of the barrier $x^*$ by pure diffusion is $\exp (-U_0/\nu_0)$; but due to 
the random growth term, this probability is multiplied by a certain 
proliferation `gain' factor $\exp {\cal G}(x^*,t)$ \cite{Rq2}. If the path $\cal C$ leading from the 
initial point $x_0$ to $x^*$ was unique, one would simply have
${\cal G}(x^*,t)=\int dt' \eta(x_{\cal C}(t'),t')$, which typically behaves as 
$\sigma_\eta \sqrt{t}$. 
In fact, 
many paths contribute to ${\cal G}(x^*,t)$. This leads to a kind of preaveraging effect of the random 
growth term $\eta$ over the width $w(t)$ of the paths $\cal C$. Therefore:
\be
{\cal G}(x^*,t) \sim \sigma_\eta \left(\int_0^t 
\frac{dt'}{w(t')^d}\right)^{1/2}.\label{Ht}
\ee
Since most paths leading to $x^*$ spend their time in the thermally 
accessible region of the well, one can estimate $w(t')$ as $w = 
L/\sqrt{U_0}$. The proliferation factor then compensates the barrier when 
$\tau \propto U_0^{3/2}$ (for $d=1$). This simple argument therefore leads to 
$b=3/2$, not very far from the numerical value $b \simeq 5/3$. Actually, one can apply this argument to the unconfined case $U_0=0$, where the detrimental 
factor is now the entropy of the random walk $\exp (-x^{*2}/t)$. Using 
self-consistently $w(t')=x^*(t')$, the compensation argument now leads to 
$x^* \propto t^{3/(4+d)}$, which is precisely the Flory result for the {\sc dp/kpz} 
problem. This Flory value can be obtained using a variational method, either with 
replicas \cite{MP}
or without replicas \cite{Orland}. In spirit, Eq. (\ref{Ht}) is actually very 
close to the 
latter calculation. The value $b=3/2$ can therefore be seen as a Flory value 
for this problem.

Returning now to the Sinai case, where the barrier height grows as $\sigma_F 
\sqrt{x^*}$, the self consistent compensation argument now leads to $\sigma_F 
\sqrt{x^*} \sim \sigma_\eta \sqrt{t/x^*}$, or $x^* \sim (\sigma_\eta/\sigma_F) 
\sqrt{t}$. The $\sqrt{t}$ behaviour is close to the numerical results shown in 
Fig. 2. However, as shown in Fig. 3, the dependence of $x_{cm}$ on $\sigma_F$
is found to be weaker than the $1/\sigma_F$ behaviour predicted by this simple argument, and closer to $1/\sigma_F^{3/4}$. We have also investigated numerically
the case where the force derives from a fractional Brownian motion with a Hurst exponent $H$. The case $H=1/2$ is the standard Sinai random walk potential considered above. An extension of the proliferation argument to this case 
predicts that $\nu^*=1/(1+2H)$ for $H>1/4$, reverting to the {\sc dp/kpz}
value $\nu^*=2/3$ for smaller values of $H$ (i.e., when the potential is
not `confining' enough). As shown in Fig. 2, our numerical values for $\nu^*$ agree quite well with this prediction: for example $\nu^*(H=3/4)\simeq 0.37$ and  $\nu^*(H=1/4)\simeq 0.65$. 

In summary, we have investigated the competition between barrier slowing down and 
proliferation induced 
superdiffusion in a model of population dynamics in a random force field. The 
one-loop {\sc rg} analysis close to the critical dimension $d_c=2$ predicts 
that the subdiffusive fixed point is unstable against `proliferation' and
flows to strong coupling. Our numerical results in $d=1$ actually suggest that both the Sinai and {\sc kpz} fixed points are unstable, and flow towards a 
new stable mixed 
fixed point. We have given a heuristic Flory like argument, which 
allows us to understand 
qualitatively the diffusive behaviour at this mixed fixed point, and also our 
results on 
proliferation assisted barrier crossing. This work can be extended in various 
directions: for 
example, a two-loop {\sc rg} calculation would be interesting. One could also study 
the effect of non linear terms in the population equation, such as $-P^2$ or 
$\vec \nabla \cdot (P \vec \nabla P)$, and the role of a non zero external force $\langle F(x) \rangle$. It would be worth performing some numerical simulations of the barrier crossing problem and of the mixed model in 
$d=2$.   

%\widetext

\noindent{Acknowledgments:} We thank M. Berg\`ere, C. de Dominicis,
T. Garel, J.M. Luck, M. Mun\~oz and K. Wiese for useful discussions.

\end{multicols}


\begin{thebibliography}{99}
\bibitem{Rq0} Note that for this choice of correlator, the force is the derivative of a random potential only in $d=1$.
\bibitem{footnoteflow} One can actually argue, using the Directed Polymer 
representation, that the {\sc dp/kpz} strong coupling fixed point is also unstable against a small static random force $F(x)$.
\bibitem{Rq2} Note that this proliferation assisted barrier crossing can only take
place if the population density $P(x,t)$ can take exponentially small values. This would not be the case, for example, if one simulates Eq. (\ref{mixed}) using a large but finite number of particles.
\end{thebibliography}
\end{document}